# Validation of Subject-Specific Knee Models from *In Vivo* Measurements

**Thor E. Andreassen[1,2*], Donald R. Hume[1], Landon D. Hamilton[1], Stormy L. Hegg[1], Sean E. Higinbotham[1], Kevin B. Shelburne[1]**

[1]Center for Orthopaedic Biomechanics, Department of Mechanical and Materials Engineering, University of Denver, Denver, CO, USA

[2]Assistive and Restorative Technology Laboratory, Department of Physical Medicine and Rehabilitation, Mayo Clinic, Rochester, MN, USA

*** Correspondence:**
Corresponding Author
thor.andreassen@du.edu

**Abstract**

Calibration to experimental data is vital when developing subject-specific models towards developing digital twins. Yet, to date, subject-specific models are largely based on cadaveric testing, as *in vivo* data to calibrate against has been difficult to obtain until recently. To support our overall goal of building subject-specific models of the living knee, we aimed to show that subject-specific computational models built and calibrated using *in vivo* measurements would have accuracy comparable to models built using *in vitro* measurements. Two knee specimens were imaged using a combination of computed tomography (CT), and surface scans. Knee laxity measurements were made with a custom apparatus used for the living knee and from a robotic knee simulator. Models of the knees were built using the CT geometry and surface scans, and then calibrated with either laxity data from the robotic knee simulator or from the knee laxity apparatus. Model performance was compared by simulation of passive flexion, knee laxity and a clinically relevant pivot shift. Performance was similar with differences during simulated anterior-posterior laxity tests of less than 2.5 mm. Additionally, model predictions of a pivot shift were similar with differences less than 3 deg or 3 mm for rotations and translations, respectively. Still, differences in the predicted ligament loads and calibrated material properties emerged, highlighting a need for methods to include ligament load as part of the underlying calibration process. Overall, the results showed that currently available methods of measuring knee laxity *in vivo* are sufficient to calibrate models comparable with existing *in vitro* techniques, and the workflows described here may provide a basis for modeling the living knee. The models, data, and code are publicly available.

## 1 Introduction

There is a widespread endeavor to create personalized modeling approaches that mimic individuals with clinically meaningful accuracy. These efforts are performed to more accurately represent the individual variability affecting functional outcomes to ultimately improve personalized medicine. Such approaches use subject-specific geometry, material properties, and loading conditions in an attempt to develop digital twins (Hassani et al., 2022; Sun et al., 2022; Viceconti et al., 2024). Investigations into the reproducibility, validity, accuracy, and limitations of personalized computational models are needed before their adoption in clinical settings (Anderson et al., 2007). Researchers have aimed to understand the effect of different model parameters (Farshidfar et al., 2022) on predictions with investigations into ligament representation and material properties (Naghibi Beidokhti et al., 2017; Peters et al., 2018), bone material properties (Peters et al., 2018; Kluess et al., 2019), and cartilage representation and material properties (Klets et al., 2016; Peters et al., 2018). Despite the documented impact of modeling decisions on model performance, the influence of the input data on the predictive abilities of models has largely been ignored. Most subject-specific models of the knee are more aptly described as specimen-specific and calibrated from measurements readily obtained from cadaveric tissue (Bloemker et al., 2015; Harris et al., 2016; Kia et al., 2016; Kluess et al., 2019; Razu et al., 2023), whereas measurements available *in vivo* to create subject-specific models of a living knee are more limited (Cooper et al., 2019).

Subject-specific models of the knee are most often constructed from medical imaging, such as computed tomography (CT) and magnetic resonance imaging (MRI). Modeled knee structures are calibrated to match material properties measured during dynamic behaviors by adjusting parameters in constitutive models; an absence of subject-specific material properties may result inpoor predictions of individual specimen kinematics (Gardiner and Weiss, 2003; Andreassen et al., 2023). As such, calibration of the material properties is necessary to improve model predictions of individual kinematics, particularly ligamentous structures at deeper flexion angles (Andreassen et al., 2023). Various subject-specific models of cadaver knees have been calibrated based on ligament forces (Kia et al., 2016; Razu et al., 2023), zero-load ligament lengths (Bloemker et al., 2015), joint distraction of the bones (Zaylor et al., 2019), or large numbers of trials of high-accuracy force-displacement measurements from robotic knee simulators (Harris et al., 2016; Chokhandre et al., 2022; Andreassen et al., 2023), all of which are impractical methodologies in living people. An implicit assumption is that the methods and data used to create models of cadaver specimens can be transferred to create models of the living knee, with no validation of this assumption having occurred.

Modeling the living knee has been constrained by limited methodology to measure *in vivo* quantities needed for soft tissue calibration. Accurate construction of living geometries is made possible using high-resolution CT, MRI, and recent statistical tools (Van Oevelen et al., 2023), but calibration of the kinetic force-displacement behavior of the knee presents a challenge without obtaining the necessary measurements. Likely for this reason, few models of a living knee exist. Tak Kang et al. and Theilen et al. each validated models built from *in vivo* data against laxity measurements (Kang et al., 2017; Theilen et al., 2023). However, like many subject-specific models in the literature (Carey et al., 2014; Shu et al., 2018), material properties were obtained from previously published values with no subject-specific calibration performed. Similarly, other researchers have used subject-specific geometries calibrated to joint laxity reported in the literature irrespective to the predicted motion of the modeled subject (Esrafilian et al., 2020). The lack of calibration to the person being modeled suggests that these models should be considered subject-specific in geometry only. Notably, Ali et al. calibrated ligament material properties to match the kinematics of passive knee flexion measured from their modeled subjects (Ali et al., 2020). Nevertheless, model predictions of loaded kinematics were not directly compared against physical measurements and instead relied on predictions from musculoskeletal modeling.

In all cases, a lack of available tools capable of making accurate, reliable, and validated kinematic and kinetic measurements of living individuals has limited the development of knee models. Fortunately, subject-specific computational modeling of the living knee may be possible with recent improvements in non-invasive measurement of knee laxity *in vivo* from the creation of several new devices (Kupper et al., 2016; Moewis et al., 2016; Pedersen et al., 2019; Andreassen et al., 2021; Shamritsky et al., 2023; Imhauser et al., 2024). These devices, and others, offer significant improvements over previous laxity measurement devices, such as the KT-1000 (Collette et al., 2012). However, the use of these non-invasive laxity tools to provide experimental measurements as targets for subject-specific knee model calibration and benchmarking has largely not been validated. Moreover, how these differences may affect model performance has not been investigated. Insights into the behavior of models built from and calibrated to *in vivo* measurements are increasingly important as subject-specific modeling becomes the basis for proposed digital twins, and *in silico* clinical trial workflows.

To support the long-term goal of creating personalized models of the living knee, we investigated whether currently available tools *in vivo* could be used to build personalized subject-specific knee models for two cadaveric knee specimens. This study aimed to validate that models calibrated to laxity measurements obtained *in vivo* are comparable to models calibrated from laxity measurements obtained *in vitro*. The results support continued work to create personalized knee models, highlighting key areas that require additional improvement. The procedures and recommendations herein can be used in future research to create *in vivo* subject-specific models with a validated dataset and process. The geometries, working models, experimental data, and code are publicly available to encourage model reproducibility.

## 2 Materials and Methods

### 2.1 Overview

To validate the use of *in vivo* laxity measurements for model creation and calibration, finite element analysis (FEA) model results were compared between two calibration methodologies: 1) FEA models calibrated to knee laxity measurements obtained using *in vivo* methods; 2) FEA models calibrated to knee laxity measurements obtained from a robotic knee simulator. Geometries of two knee specimens (Table 1) were obtained from a combination of lower-extremity CT scans, and surface scans of the bones and soft tissues. Measurements for calibration were obtained using a previously validated knee laxity apparatus (KLA) (Andreassen et al., 2021) designed to measure knee laxity *in vivo* and a robotic knee joint simulator (RKS). *In vivo* laxity experimentation, was recorded first using two intact lower body cadaveric specimens using the KLA, and then subsequent dissection performed to facilitate RKS testing on the same specimens.

The CT and surface scans were used to create model geometries for both specimens. These models were then calibrated against two different laxity datasets. In one case, models were calibrated to laxity measurements from the knee laxity apparatus, known as the “KLA” models. The other case was calibrated to laxity measurements from the robotic knee simulator, known as the “RKS” models (Figure 1). The two models were then used to predict anterior-posterior laxity at various knee flexion angles and a passive knee flexion. Kinematics and ligament force predictions were compared. Additionally, models were used to predict a simulated pivot shift and resulting kinematics and ligament loads compared.

The knee modeling process followed the Team DU workflow from the KneeHub project (SimTK: Reproducibility in Simulation-Based Prediction of Natural Knee Mechanics: Project Home, n.d.; Erdemir et al., 2019; Rooks et al., 2021; Andreassen et al., 2023) to allow for a comparison with previous work that examined the differences in modeling strategy. The following sections describe the experimental and modeling workflow.

### 2.2 Experimental Data Collection

The experimental data was collected as part of previous work (Andreassen et al., 2021) and is summarized herein. Two non-frozen male pelvis-to-toes cadavers (Table 1) were obtained with no history of musculoskeletal ailments. Prior to testing, specimens underwent CT scans (Figure 1). CT scans were collected axially (Siemens SOMATOM Perspective, Erlangen, Germany) with approximately 0.75 mm x 0.75 mm in-plane resolution and a 0.6 mm axial resolution from approximately L5 to the toes of both legs. Bone geometries were segmented from CT scans using a combination of global thresholding and manual segmentation methods and exported as STLs (Simpleware ScanIP, Synopsys, Sunnyvale, CA).

Specimens were placed in a custom knee laxity apparatus (KLA) (Andreassen et al., 2021) designed to measure laxity in the living knee. To simulate a standard knee laxity data collection with a living subject, a series of loads were applied to anterior, internal, and external degrees of freedom (DOF) of the tibia at 30 and 90 deg of knee flexion. Maximum loads were approximately 175 N for anterior and 5.5 N*m for internal and external as measured via load cell. To approximate a passive knee extension while stereo radiography images were recorded, a cuff was placed around the ankle and attached with a cable and rod to manually push the knee to deep flexion (~150 deg) and pull to full extension. This simulated passive knee extension is later referred to as the experimental "*KLA experiment*" kinematics. The resulting displacements for all DOF were recorded using 3D image tracking techniques from high-speed stereo radiography (HSSR) images (Ivester et al., 2015; Kefala et al., 2017).

Immediately following these measurements, specimens were dissected, leaving approximately 230 mm of soft tissue and bone intact above the knee joint line and 200 mm below the knee joint line. Each knee specimen was cemented into custom femur and tibia-fibula fixtures and affixed to a VIVO robotic knee simulator (RKS) (AMTI, Watertown, MA). Additionally, a custom quadriceps actuator (Behnam et al., 2024) was affixed to the quadriceps tendon to simulate the passive tension in the quadriceps tendon (McKay et al., 2010). The joint simulator applied laxity loads for anterior-posterior (AP), internal-external (IE), and varus-valgus (VrVl) between 0-120 deg of knee flexion in 15-degree increments. Maximum loads applied were approximately 200 N for AP, 7.5 N*m for IE, and 10 N*m for VrVl and measured using a built-in 6 DOF load cell. The resulting displacements for all DOF were recorded using an Optotrak motion capture system (NDI, Ontario, Canada). Following laxity testing, the passive range of motion of the knee was recorded in the simulator with no loads applied and is later referred to as the experimental "*RKS experiment*" kinematics.

The laxity values used during the model calibration (described below) were selected as a subset of the overall measurements collected from experimentation and differed between the RKS and KLA data sources because of experimental constraints and the model optimization procedures. From the KLA data, laxity values used for model calibration were the anterior and IE knee laxity at 30 and 90 deg of knee flexion at various load levels. From the RKS data, laxity values used for model calibration were the AP, IE, and VrVl laxity at 0, 30, 60, and 90 deg of knee flexion at the maximum and minimum loads measured. Due to experimental limitations, the target used was chosen at 75 instead of 90 deg of knee flexion in some cases. In all models, an additional calibration target was placed at 0 deg of knee flexion taken from the kinematics of the passive range of motion at full extension.

The knee specimens were further dissected after experimentation, leaving the bones, ligaments, and knee capsule intact. A white-light scanner (Artec Space Spider, Artec, Santa Clara, CA) was used to scan the surface of the knees (Figure 2). Then, the soft-tissue structures were removed, and ligament attachment sites of the major knee ligaments (ACL, LCL, MCL, PCL) and tendons (patellar tendon) were outlined on the bones, and the bones were scanned again with the scanner (Figure 2). This process resulted in surface with color texture scans of the full intact knee capsule as well as individual separate scans for the femur, tibia-fibula, and patella (Figure 2). Fiducial screws and stickers were added to the bones prior to scanning to allow for easier registration, but did not affect the experimental kinematics measurements. The resulting geometries were exported as STLs.

Geometries from the CT scans were used to create local bone coordinate systems in the Transepicondylar (TEA) axis coordinate system (Figure 3) following the joint coordinate system convention from Grood and Suntay (Grood and Suntay, 1983). All kinematics from the KLA and the RKS testing were represented in the same local coordinate system of the bones.

### 2.3 Geometry Identification

Geometries of the bones were created from a combination of CT scans and surface scans. Geometries of the bones from the surface scan were first aligned to the position of the bones in the CT scan with a combination of manual and automatic registration using an iterative closest point (ICP) algorithm in MATLAB (Mathworks, Natick, MA). Final model geometries of the bones were created by cropping the bones obtained from the CT images to the region around the knee (approximately 150 mm above and below the joint line). Geometries of cartilage were obtained by Boolean subtraction of the aligned CT scans of the bones with the corresponding surface scans of the bones and cartilage. In all cases, geometries were smoothed and fixed (removal of poor-quality elements, removal of inaccuracies from segmentations, etc.) using a combination of MeshMixer (Autodesk, San Francisco, CA) and MeshLab (Cignoni et al., 2008). Marked attachment site regions from the surface scans for the major ligaments (ACL, LCL, MCL, PCL) and patellar tendon were projected along surface normal directions to the CT bones to determine the approximate attachment on the true boney surface. For ligaments that could not be easily identified during experimentation, approximate attachment sites were identified using descriptions from the literature, summarized in Table 2 (LaPrade et al., 2003, 2007, 2021; De Maeseneer et al., 2004; Petersen and Zantop, 2007; Liu et al., 2010; Claes et al., 2013; Chahla et al., 2020).

## 2.4 FEA Model Development

### 2.4.1 Overview

FEA models of the knee were created in Abaqus Explicit (Dassault Systemes, France) using a previously described modeling workflow (Rooks et al., 2021) (Figure 3 and Figure 4). Models of each specimen were created in the initial position of the bones defined by their full-extension position in the CT. Rigid body reference nodes were defined for the femur and tibia/fibula geometries. Cylindrical joints were created between the femur and tibia/fibula rigid body nodes following the Denavit-Hartenberg convention described by Grood and Suntay (GS) (Grood and Suntay, 1983). These joints allowed the application of loads or displacements to each degree of freedom (DOF) along the cylindrical joints for medial-lateral (ML), anterior-posterior (AP), and superior-inferior (SI) or as torques or rotations around the cylindrical joints for flexion-extension (FE), varus-valgus (VrVl), and internal-external (IE). The bony surface of the cartilage geometry was rigidly fixed to the rigid body nodes for the femoral and tibial cartilage. Ligaments were defined with 1D connectors and rigidly fixed to the bones. Contact was modeled between the cartilage surfaces of the tibia and the femur.

### 2.4.2 Bone Meshes

Bones of the femur and a combined tibia and fibula were modeled as rigid triangular surfaces (R3D3 elements in Abaqus) rigidly fixed to a rigid body node following previous work (Harris et al., 2016; Rooks et al., 2021). Elements of geometries were approximately 1.5 mm in size.

### 2.4.3 Cartilage Meshes

Cartilage geometries generated for the distal femur and the medial and lateral proximal tibia were exported to Hypermesh (Altair, Troy, MI), and the articular surface and bony surfaces were identified. A quadrilateral mesh was created on the bony surface. The quadrilateral mesh and the original articular and bony surfaces were used in Hypermesh to create 3D reduced-integration hexahedral cartilage meshes (C3D8R elements in Abaqus). While stress was not critical to this study, the cartilage was meshed as hexahedral elements with appropriate element sizes to allow for future use for stress analysis with non-linear material models. Following previous methods (Halloran et al., 2005; Fitzpatrick et al., 2010; Huff et al., 2020), the cartilage was modeled with a calibrated (unique to each specimen) tri-linear pressure overclosure material property to improve computational performance. The details of this calibration are provided in the Supplemental Material.

A mesh convergence study was performed following recommendations for calculation verification from Anderson et al. (Anderson et al., 2007) and reported in the Supplemental Material. All cartilage geometries had target element lengths at the largest between 1 mm and 0.5 mm.

After the hexahedral meshes were created from the original triangulated surfaces, initial overclosures were observed between the femoral and tibial cartilage. Overclosures often create instability and convergence problems for explicit finite element analysis (FEA) models. Our previously developed and publicly available code package and corresponding algorithm using generalized regression neural networks (GRNNs) was used to remove initial overclosures between cartilage geometries via equal weighting of the resulting deformations between tibial and femoral cartilage (Andreassen et al., 2024).

#### 2.4.4 Ligament Connectors

Ligaments were modeled as non-linear 1D tension-only spring connectors (Axial type CONN3D2 elements in Abaqus) like those previously used (Blankevoort and Huiskes, 1996) and formally described by Yu et al. (Yu et al., 2001). Ligaments were modeled with a reference strain parameter defining the initial tension present in the ligament in its initial configuration and the stiffness of the ligament in the linear region (Table 3). The range of values used for the reference strain and ligament stiffness were approximately the same as those reported in the "Knee Model Calibration Specification" document for Team DU in the KneeHub Project (SimTK: Reproducibility in Simulation-Based Prediction of Natural Knee Mechanics: Project Home, n.d.; Erdemir et al., 2019; Rooks et al., 2021; Andreassen et al., 2023). A constant quadratic toe-in region was created for all ligaments with an assumed strain parameter of 0.03 (Blankevoort et al., 1991). Ligaments were separated into models of individual bundles based on anatomical descriptions, with several fibers modeled for each bundle. The location of the individual fibers was determined by visually identifying the approximate major axis of the ligament attachment region and choosing points equidistant along the major axis based on the number of desired fibers to approximate the span of the overall region (Supplemental Material). Reference strain and stiffness parameters were unique for each ligament, but all fibers within a single bundle shared the same material properties. Ligament attachments were tied to the respective bone's rigid body nodes (multi-point constraint beam type in Abaqus).

#### 2.4.5 Simulation of Knee Motion

The position of the tibia and fibula rigid body nodes was rigidly fixed (boundary encastre in Abaqus), while the position of the femur was determined by the joint connectors. All DOF for the joint connectors were placed in load control except for the FE connector, which applied the desired knee flexion angle in displacement control. All simulations of knee dynamics utilized two sequential steps in Abaqus Explicit.

The first step, a settling step, began with the bones in their initial CT full-extension pose, and applied a desired compression level in the SI direction to the cylindrical SI connector (connector load in Abaqus) using a load starting at 0 N and linearly ramping to the desired compression level. The model was highly damped in the first step to reduce vibrations caused by ligament tension as the model settled into a stable initial pose. Damping was applied to the joint connectors (connector damping in Abaqus) for the translation and rotational DOF. To ensure the high damping did not affect the motion in the second step, the damping was defined as dependent on temperature. A high temperature during the first step resulted in significant damping (100 Ns/mm for translation and 100 N*s*mm/rad for rotation).

During the second step, the knee was flexed to the desired knee angle by rotating about the cylindrical FE connector (connector motion in Abaqus), and target loads were applied to the respective DOF (connector load and CLOAD in Abaqus). This step had a low temperature resulting in negligible damping. The load in each DOF was linearly increased to the target value and held constant for the final 30% of the step.

### 2.5 Model Calibration

Ligament reference strain and stiffness parameters were calibrated in an optimization process that simulated knee model movement in response to target loading conditions and minimized the error between the measured and predicted kinematics (Andreassen et al., 2023). Specifically, a set of calibration targets (matched kinematics and load data) was defined from the laxity measurements described above at discrete knee flexion angles and levels of applied load. Separate calibration targets were created from the laxity measurements made using the KLA and the RKS (Figure 5).

The optimization process was managed in MATLAB. For each iteration of the optimization process, a custom MATLAB script was used to set the ligament parameters (26 parameters total, Table 3) in Abaqus for the simulation of a given calibration target. Using the Abaqus API, custom Python scripts extracted simulation results, including kinematics. The squared error between simulated GS kinematics and experimental GS kinematics was calculated for each kinematic DOF. This was repeated for each of the given calibration targets to calculate the optimization cost function (described below). The optimization process occurred in two phases for each knee model. First, a particle swarm global optimization (Kennedy and Eberhart, 1995) narrowed the search space to the location of the most likely global minimum. Then, using the ligament parameters at this approximate minimum as an initial point, a Nelder-Meade Simplex solver determined the true local minimum around this point. The approximate number of iterations to reach a minimum was 750 for the particle swarm optimization and 500 for the Nelder-Meade Simplex solver (Nelder and Mead, 1965). While the exact time required to complete an iteration for each model calibration depended on the number of calibration targets and elements within the model, the average clock time was approximately 210 seconds per iteration. Therefore, the overall time to complete calibration for each knee model was approximately 73 hours (single Intel Xeon Gold 6134 CPU @ 3.2 GHz).

The optimization process minimized a cost function consisting of the squared error between the measured and simulated calibration targets and a penalty term. Trials were grouped in similar categories (e.g., anterior laxity at 30 deg of knee flexion for multiple loads), and the $75^{th}$ percentile of the root mean squared error (RMSE) for each DOF of a given group was calculated and normalized to the range between the minimum and maximum observed for each kinematic DOF across all experimental results (Figure 6). Targets were grouped together to bias the optimization across the range of flexion angles and DOF (rather than, for instance, AP at 30 deg at 10 N, 20 N, 30 N and 40N). The normalized errors for each DOF were then scaled by chosen scalar weights and summed across all categories into a total cost. This weighting allowed selected DOF to be more emphasized based on the primary DOF for a given laxity trial (AP for anterior at 30 deg, IE for internal at 90 deg, etc.) while allowing for secondary DOF (IE for anterior at 30 deg, AP for internal at 90 deg, etc.) to also be included with less emphasis. Additionally, a penalty term squared the cost if any trials reached joint limits on the SI and ML DOF. The penalty improved optimization speed by quickly guiding the search away from unrealistic solutions. An example calculation from a single iteration of the optimization for the Specimen 2 model calibrated to KLA targets is included in the Supplementary Material as a spreadsheet file.

## 2.6 Model Performance Comparison

### 2.6.1 Ligament Parameters

The resulting ligament material property parameters (reference strain and stiffness) for each calibrated model were compared to evaluate the differences between the KLA and RKS models, and between specimens. Values reported are the true parameters calibrated for each model. However, because of the large range of resulting material parameters, particularly between reference strains and stiffnesses, percentage differences were calculated in each case to simplify model and specimen comparisons.

### 2.6.2 AP Laxity

To compare AP laxity of the models, AP loads of 133N and -133N were applied to each model at 30, 60, and 90 deg of knee flexion. A 133N (30 lbf) load was chosen as it is commonly used when evaluating knee laxity clinically (Un et al., 2001; Starkel et al., 2014). Simulation of AP laxity was performed as described above with the addition of a third simulation step in Abaqus/Explicit that linearly increased and decreased the AP loading between 133N and -133N. The GS kinematics for the AP direction were recorded and root-mean-squared difference (RMSD) was calculated between the models calibrated with laxity measurements from the KLA and RKS.

### 2.6.3 Passive Flexion

Passive flexion was simulated by applying zero loads for all DOF and prescribing knee flexion angle from 0 to 90 deg. A two-step Abaqus/Explicit simulation was performed as described above. Results were obtained for the predicted passive flexion kinematics of AP, IE, and VrVl vs. knee flexion angle. Results were compared to the experimental values measured for passive flexion with the KLA and RKS.

### 2.6.4 Pivot Shift

To compare the models during complex motions, a simulated pivot shift test was performed. The pivot shift test is a clinical evaluation that aims to determine the stability of the knee as a means of predicting possible anterior cruciate ligament (ACL) injury (Matsushita et al., 2013). Following previous work, a pivot shift was simulated by placing the knee in 30 deg of flexion and applying an 8 N*m valgus torque combined with a simultaneous 4 N*m internal torque (Schafer et al., 2016; Thein et al., 2016). A two-step simulation was performed as described above. Model kinematics were recorded for the KLA and RKS models for both specimens. In addition, an ACL-deficient pivot shift was simulated for all models and calibrations. The ACL-deficient condition was simulated by creating a parameter that controlled the presence of a failure in the connectors representing the ACL for both the anteromedial and posterolateral bundles (connector failure in Abaqus). A parameter equation was created that caused an immediate failure of the ACL connectors in the first simulated time increment of the model. Following the failure of the connectors, the simulation progressed through the remaining time steps as if the ACL was not present. Kinematics and ligament forces were compared between models for the intact and ACL-deficient conditions.

# 3 Results

## 3.1 Ligament Parameters

The calibrated material properties for ligament reference strain and stiffness are reported in Table 4. The smallest reference strain on average was observed in the dMCL with a value of 0.85 and the largest observed in the PCAP_M with a value of 1.17. The smallest stiffness on average was observed in the PFL with a value of 51.9 N/mm and the largest observed in the dMCL with a value of 138.9 N/mm. Percent differences between stiffnesses were approximately 4 times greater than the percent differences observed for reference strains on average. The percent differences for the reference strains were lower between models (KLA vs. RKS) as compared with between specimens, with the average differences found of 4.6% and 12.3% for inter-model (KLA vs. RKS) and inter-specimen (Specimen 1 vs. Specimen 2), respectively. The same effect was observed for stiffness where the average difference was 14.3% and 55.8% for inter-model and inter-specimen, respectively.

## 3.2 AP Laxity

Predicted AP translation in response to 133N anterior and posterior load was similar between the models calibrated with KLA and RKS measurements (Figure 7). The RMSD in the anterior direction between the KLA and RKS models was 2.38 mm and 1.94 mm for Specimen 1 and Specimen 2, respectively. RMSD in the posterior direction was 2.45 mm and 1.90 mm. The ACL was the most loaded ligament for the anterior laxity trials at all flexion angles for all models (Table 5). For the posterior direction, the PFL and POL were loaded for both models in Specimen 1. The PCL was the most loaded ligament for Specimen 2. The RMSD of ligament loads between the KLA and RKS models was 10.3 N and 11.4 N for Specimen 1 and 2, respectively.

## 3.3 Passive Flexion

Predicted AP translation of the tibia was similar in magnitude and trend for both models and specimens during simulated passive flexion (Figure 8). While the prediction of IE and VV was similar for both models and specimens, some differences were noted. The RMSD between the KLA and the RKS model was 3.5 mm, 2.6 deg, and 0.4 deg for AP, IE, and VrVl, respectively, for Specimen 1 and 1.1 mm, 1.2 deg, and 0.3 deg for Specimen 2.

### 3.4 Pivot Shift

Kinematics during the simulated pivot shift were within 2.6 deg and 2.8 mm between the KLA and RKS models for all simulations for rotations and translations, respectively (Table 6). The simulated ACL-deficient condition for Specimen 2 was an exception, where both the RKS and KLA models predicted a dislocation. All models predicted increases in tibial anterior translation and valgus rotation for the ACL-deficient condition relative to the intact model. All models predicted an increase in the anterolateral structure (ALS) ligament load for the ACL-deficient condition compared with the intact condition (Table 7). Ligament loads for the collateral ligaments (LCL and MCL) were zero for both models of Specimen 1 but were non-zero for Specimen 2. In both the KLA and RKS models for Specimen 2, the sum of the LCL and MCL ligament loads decreased for the ACL-deficient condition relative to the intact condition, with the KLA model decreasing by 67.4 N, while the RKS model decreased by 53.2N.

## 4 Discussion

Recent calls for more personalized approaches to medicine, including Digital Twins and *in silico* clinical trials, have prompted an increased demand for computational models of living people. Personalized knee models could be used in conjunction with existing surgical planning tools to better predict and understand the short- and long-term outcomes of various treatment options. However, the necessary tools to obtain the measurements of the living knee are limited. While knee laxity is a routine clinical evaluation, these measurements have historically been insufficient to calibrate models with useful accuracy. Furthermore, while recent work has examined the effects of modeling methodologies on model performance, the impact of the data used to build and calibrate models has received little attention. This study investigated the effects of experimental inputs on model predictions following model calibration using measurements from *in vivo* (KLA) and *in vitro* (RKS) methods. Specimen-specific FEA models of the knee were developed and calibrated, and model performance was compared. Our results showed that accurate model calibration can be achieved using measurements available to living subjects.

Whether calibrated from KLA or RKS measurements, the models captured the distinctly different behavior of the two knee specimens. The differences in calibrated material parameters were greater inter-specimen as compared to inter-model (Table 4). This highlights that the models built using different calibration data (KLA vs RKS) can capture the unique material behavior of each specimen. However, for almost every ligament, we observed that the percent differences for reference strain were smaller than stiffness. Previous work from Baldwin et al. used Monte Carlo and Advanced Mean Value (Wu et al., 1989) analyses to determine the relative importance of model parameters on knee joint laxity (Baldwin et al., 2009). They found that the reference strain was frequently more important than stiffness for accurate recreations of joint laxity. This may explain why reference strain values in our models were more similar than the overall ligament stiffness, because accurate recreation of joint motion, particularly in loaded conditions, was more sensitive to reference strain rather than stiffness.

For each specimen, results were similar for both the passive flexion and the AP laxity simulations despite the differing calibration targets from each data source (Figure 5, and Figure 6). The RKS measurements provided laxity data for 3 DOF at four knee flexion angles while the KLA *in vivo* measurement device provided laxity in only 2 DOF at two knee flexion angles. Still, the predictions of the AP displacement in response to applied load were similar and unique to each specimen, with similar ligament loading observed for all AP conditions (Table 5). The resulting errors between model predicted kinematics for the AP conditions across both specimens were less than 2.5 mm and 2.0 mm for anterior and posterior laxity, respectively (Figure 7); these errors are within the minimum detectable change (MDC) for AP laxity reported from other *in vivo* knee laxity devices, with MDCs ranging from 1.1 mm to 4.5 mm (Mouton et al., 2015; Smith et al., 2022; Imhauser et al., 2024). Notably, posterior knee laxity was accurate despite no posterior loading targets in the KLA model calibration. Additionally, model predictions of AP laxity were similar at 60 deg of knee flexion, despite the KLA model not having this flexion angle in the calibration targets for either specimen. Furthermore, the low RMSD between KLA and RKS models during passive flexion (flexion free of dynamic loads) suggests that both models predict nearly the same kinematics despite not being calibrated to passive data. For both specimens, differences between calibration with RKS or KLA data was within the errors reported for passive flexion from another study, wherein knee models were built from the same experimental data but with different modeling workflows (Andreassen et al., 2023). These results demonstrate that laxity measurements from *in vivo* techniques, such as those from the KLA, can provide sufficient targets for model calibration in subject-specific modeling.

The models calibrated from RKS and KLA predicted similar kinematics for a simulated pivot shift test. The maximum differences between model predictions were less than 3 deg and 3 mm for rotation and translation, respectively. Inter-specimen differences in kinematics were far greater than inter-model differences in both ACL-intact and deficient conditions (Table 6). Additionally, both the KLA and the RKS models predicted the same dislocation behavior in Specimen 2 for the ACL-deficient condition. In agreement with Thein et al., force in the ALS of the knee increased without the ACL in all models (Thein et al., 2016). This demonstrated that models calibrated using data acquired with *in vivo* methods can make meaningful predictions beyond the calibration data, including dislocation behavior. Still, while similar ligament loads were observed between the KLA and RKS models for the ACL-intact and deficient conditions, differences remain. For Specimen 2, in the intact condition with KLA calibration, the LCL force was 0.0 N and the superficial MCL force was 215.4 N; in contrast, the RKS calibration predicted the LCL force to be 190.2 N and the superficial MCL force to be 88.6 N (Table 7). These results highlight that while models may yield similar joint-level force-displacement behavior from different ligament material properties, the resulting ligament loads may be variable. Similarly, recent work by Theodorakos et al. showed that model calibration is sensitive to initial conditions for the material properties used for model calibration (Theodorakos and Andersen, 2024). They showed that different initial conditions resulted in different material properties following model calibration and different ligament forces despite small overall kinematic and kinetic differences at the joint-level. Thus, subject-specific models created to predict *in vivo* ligament loads may not provide accurate results using joint-level calibration alone. Additional calibration constraints informed by subject-specific information, such as penalties on ligament loads, or a narrowing of the range of possible ligament parameter values prior to model calibration, may be necessary to drive the calibration to a set of material properties that ensure feasible predictions of ligament loads.

Even so, modeling workflow may be a more important factor than the source of the calibration data. The aforementioned KneeHub project had 5 different groups develop, calibrate, and benchmark two specimen-specific models using individual strategies but with the same data (Erdemir et al., 2019; Rooks et al., 2021; Andreassen et al., 2023). Using supplementary data reported from that work, average inter-model RMSD in passive flexion were as high as 6.2 mm, 14.9 deg, and 6.8 deg for AP, IE, and VrVl, respectively; considerably higher than the maximum values found herein of 3.5 mm, 2.6 deg, and 0.4 deg, for AP, IE, and VrVl, respectively. Moreover, an average 10% difference in inter-model reference strain was found between the 5 modeling strategies, in that work, compared with the 4.6% inter-model observed herein. These differences suggest that given a similar level of accuracy between measurement methodologies, the method by which it is collected, and the specific targets used for calibration, may matter less than the choice of modeling workflow.

This study had limitations. The first limitation is the small number of knee specimens utilized, which limited the power of the study to investigate subject variation. Even though the sample size of two specimens was comparable to other studies of subject-specific knee modeling (Kia et al., 2016; Ali et al., 2017; Razu et al., 2023), future work should include larger sample sizes and specimen variation. Another limitation is the applicability of the laxity measurements made herein to those performed in living individuals. In measuring knee laxity *in vivo*, there is the potential for physiological factors such as passive muscle tone, coactivation, spinal reflexes, pathology, and other contributions to muscle force that influence the amount of knee laxity measured from the passive structures alone. Previous work has shown that laxity in the knee during an anterior drawer test increases in patients under anesthesia compared to when awake (Matsushita et al., 2013). Future work using living individuals should include methods to reduce the possibility of muscle-reduced knee laxity by employing muscle stretch-relaxation techniques (Osternig et al., 1987) or fatiguing muscle contractions (Nawata et al., 1999), which has been shown to increase knee laxity. In addition, future studies involving living subjects should include methods to determine the relative activation of muscles such as electromyography (EMG). The final limitations are the choice of representation of cartilage as linear elastic isotropic and the lack of a meniscus model. The cartilage model was simplified to decrease computational burden and previous work has shown that models utilizing linear-elastic isotropic representations of cartilage can accurately predict experimentally measured joint contact (Kiapour et al., 2014). Furthermore, as this work aimed to examine the kinematics of the knee and not contact patterns or stress, the choice of cartilage material property likely had little effect on the observed results. Still, future work should investigate if there are significant effects of the choice of cartilage material models on ligament material calibration. The models did not include the menisci, in line with other studies (Farshidfar et al., 2022), as inclusion of the menisci has been shown to have little effect on the kinematics of the knee at less than 90 deg of knee flexion (Amiri et al., 2006). Still, for certain contexts of use, inclusion of menisci is crucial. Future work should investigate if the presence of menisci influences ligament material calibration.

In summary, this study reported small errors between the models calibrated to data from a laxity measurement apparatus, designed for the living knee, compared with models calibrated to data from a robotic knee joint simulator. The viability of using knee laxity measurements in future calibration of living subjects was demonstrated by close agreement with knee calibration using measurements from cadaveric testing. The workflows and optimization strategies described here act as a basis for future subject-specific modeling and the development of digital twins. The models, results, and tools created are publicly available to encourage model reproducibility.

## 5 Conflict of Interest

The authors declare that the research was conducted in the absence of any commercial or financial relationships that could be construed as a potential conflict of interest.

## 6 Author Contributions

TEA collected experimentation data, processed original data, created models, ran model calibration, performed results analysis, and wrote the initial draft of the manuscript. DRH helped collect experimentation data and created the initial model framework. LDH, SLH, and SEH collected experimentation data and processed the original data. KBS originated the idea for the study, collected experimentation data, reviewed results, and wrote the initial draft. All authors reviewed and edited the final manuscript.

## 7 Funding

NIH National Institute of Arthritis and Musculoskeletal and Skin Diseases, National Institute of Biomedical Imaging and Bioengineering, and the Eunice Kennedy Shriver National Institute of Child Health and Human Development through the following grants: U01 AR072989 and T32 AR056950.

## 8 Acknowledgments

The authors would like to thank Dr. Charlie Ho at the University of Colorado School of Medicine for his help in verifying the accuracy of segmented structures. The authors would also like to thank Gary Doan, Rachel Wathen, Yashar Behnam, and Chadd Clary at the University of Denver for their help with the original cadaveric experimentation that provided the data used herein.

## 9 Data Availability Statement

Specimen imaging, dynamics measurements, working Abaqus models, calibration code, post-processing code, and results are publicly available online at the following repositories:

https://simtk.org/projects/in_vivo_valid

https://zenodo.org/records/10416664

https://zenodo.org/records/10521352

https://datadryad.org/stash/dataset/doi:10.5061/dryad.zkh1893gw

TABLES

Table 1: Donor specifics for knee models. Specimen IDs are used for Supplemental Material, including model files, where the data is referred to using the Specimen ID rather than Specimen 1 and Specimen 2.

| | Specimen 1 | Specimen 2 |
|---|---|---|
| Specimen ID | S192803 | S193761 |
| Modeled Side | L | L |
| Sex | M | M |
| Age (years) | 29 | 64 |
| Height (cm) | 188 | 178 |
| Weight (kg) | 113.4 | 56.2 |
| BMI ($kg/m^2$) | 32.1 | 17.8 |

Table 2: Modeled knee ligaments and anatomical descriptions of attachment sites and any changes used.

| Ligament | Ligament Major Group | Ligament Abbreviation | Literature Description of Ligament Origin | Literature Description of Ligament Insertion | Adjustments to Attachment Sites |
|---|---|---|---|---|---|
| Anteromedial Bundle of ACL | ACL | ACL_AM | Posterior portion of the lateral femoral condyle. Posterior to lateral intercondylar ridge. Superior to bifurcate ridge(Petersen and Zantop, 2007). | Centralized in the ML direction on the tibial plateau, at approximately 30% of the total AP length of the tibia from the anterior side(Petersen and Zantop, 2007). | None |
| Posterolateral Bundle of ACL | ACL | ACL_PL | Posterior portion of the lateral femoral condyle. Posterior to lateral intercondylar ridge. Inferior to bifurcate ridge(Petersen and Zantop, 2007). | Centralized in the ML direction on the tibial plateau, at approximately 44% of the total AP length of the tibia from the anterior side(Petersen and Zantop, 2007). | None |
| Main Bundle of LCL | LCL | LCL | Approximately 1.4mm superior and 3.1 mm posterior to the lateral epicondyle of the femur(LaPrade et al., 2003). | Inserts into fibula head approximately 8mm posterior of the anterior portion of the fibular head, and approximately 28 mm distal to the fibular head apex(LaPrade et al., 2003). | None |
| Superficial Anterior Fiber of MCL | MCL | MCL_SA | Slightly superior and anterior to the medial epicondyle of the femur(Liu et al., 2010). | Anterior region of the medial side of the tibia approximately 6 cm distal to the tibial joint line. Additional insertion around the most medial portion of the tibial plateau(Liu et al., 2010). | In cases with a rapidly narrowing tibia, insertion was chosen to be the proximal attachment of the superficial MCL, rather than the distal one to approximate correct line of action. |
| Superficial Middle Fiber of MCL | MCL | MCL_SM | Slightly superior to the medial epicondyle of the femur(Liu et al., 2010). | Middle region of the medial side of the tibia approximately 6 cm distal to the tibial joint line. Additional insertion around the most medial portion of the tibial plateau(Liu et al., 2010). | In cases with a rapidly narrowing tibia, insertion was chosen to be the proximal attachment of the superficial MCL, rather than the distal one to approximate correct line of action. |
| Superficial Posterior Fiber of MCL | MCL | MCL_SP | Slightly superior and posterior to the medial epicondyle of the femur(Liu et al., 2010). | Posterior region of the medial side of the tibia approximately 6 cm distal to the tibial joint line. Additional insertion around the most medial portion of the tibial plateau(Liu et al., 2010). | In cases with a rapidly narrowing tibia, insertion was chosen to be the proximal attachment of the superficial MCL, rather than the distal one to approximate correct line of action. |
| Deep Bundle Fiber of MCL | MCL | MCL_D | Posterior and inferior to the superficial MCL and the medial epicondyle(Liu et al., 2010). | Medial aspect of the tibial plateau approximately 6mm from the tibial joint line(Liu et al., 2010). | None |
| Anterolateral Bundle of PCL | PCL | PCL_AL | Located on the anterior side of the medial condyle in the intercondylar fossa. Inferior to the medial intercondylar ridge. Anterior to the medial arch point(Chahla et al., 2020; LaPrade et al., 2021). | Anterolateral to PCL_PM and near the posterior medial edge of the lateral meniscus(Chahla et al., 2020; LaPrade et al., 2021). | None |
| Posteromedial Bundle of PCL | PCL | PCL_PM | Located on the anterior side of the medial condyle in the intercondylar fossa. Inferior to the medial intercondylar ridge. Posterior to the medial arch point(Chahla et al., 2020; LaPrade et al., 2021). | Edge of the champagne glass dropoff (CGD) of the tibial plateau in the intercondylar facet(Chahla et al., 2020; LaPrade et al., 2021). | None |
| Anterolateral Structure | ALS | ALS | Originate on the lateral epicondyle of the femur just anteriorly to the origin of the LCL. In many cases, the origin is more superior and joins with the portion of the LCL(Claes et al., 2013). | Posterior to Gerdy's tubercle on the tibial plateau. Approximately found at the intersection of a ray cast between the Gerdy's Tubercle and the fibular head(Claes et al., 2013). | None |
| Popliteofibular Ligament | PFL | PFL | Approximately 18.5mm from the LCL origin on the femur in the inferior/anterior direction(LaPrade et al., 2003). | Inserts into the fibula approximately 3mm inferior to the apex of the fibula head on the anteromedial slope(LaPrade et al., 2003). | PFL femoral attachment was moved approximately to the medial epicondyle to approximate the line of action of the force of the combined popliteofibular ligament and popliteal tendon rather than true anatomic accuracy. |
| Posterior Oblique Ligament | POL | POL | Superior and posterior to the origin of the Superficial MCL. Approximately 8 mm inferior and 6.4 mm posterior to the adductor tubercle and 1mm inferior and 3mm anterior to the gastrocnemius tubercle(LaPrade et al., 2007). | Slightly inferior to the tibial posteromedial portion of the tibial plateau near the posteromedial portion of the medial meniscus(LaPrade et al., 2007). | None |
| Medial Posterior Capsule | PCAP | PCAP_M | Originates on the posteromedial portion of the femoral cortex a few centimeters above the most superior portion of the femoral cartilage(De Maeseneer et al., 2004). | Posteromedial portion of the tibial plateau, approximately 1-2 cm below the knee joint line(De Maeseneer et al., 2004). | Femoral origin was moved to the edge of the posterior condyles of the femur, where the contact of the capsule with the condyles would occur. |
| Lateral Posterior Capsule | PCAP | PCAP_L | Originates on the posterolateral portion of the femoral cortex a few centimeters above the most superior portion of the femoral cartilage(De Maeseneer et al., 2004). | Posterolateral portion of the tibial plateau, approximately 1-2 cm below the knee joint line. | Femoral origin was moved to the edge of the posterior condyles of the femur, where the contact of the capsule with the condyles would occur. |

Table 3: Modeled ligaments organized by bundle, number of modeled fibers, and design variable assigned to ligament material parameter. $X_1$ represents the first design variable, $X_2$ represents the second design variable, and so on.

| Ligament | Ligament Major Group | Ligament Abbreviation | Number of Fibers | Reference Strain Design Variable | Reference Strain Range | Stiffness Design Variable | Stiffness Range (N/mm) |
|---|---|---|---|---|---|---|---|
| Anteromedial Bundle of ACL | ACL | ACL_AM | 2 | $X_1$ | [0.95-1.25] | $X_{15}$ | [50-150] |
| Posterolateral Bundle of ACL | ACL | ACL_PL | 2 | $X_2$ | [0.95-1.25] | $X_{16}$ | [50-150] |
| Main Bundle of LCL | LCL | LCL | 3 | $X_3$ | [0.55-1.15] | $X_{17}$ | [60-200] |
| Superficial Anterior Fiber of MCL | sMCL | sMCL_A | 1 | $X_4$ | [0.70-1.05] | $X_{18}$ | [40-180] |
| Superficial Middle Fiber of MCL | sMCL | sMCL_M | 1 | $X_5$ | [0.70-1.05] | $X_{18}$ | [40-180] |
| Superficial Posterior Fiber of MCL | sMCL | sMCL_P | 1 | $X_6$ | [0.70-1.05] | $X_{18}$ | [40-180] |
| Deep Bundle Fiber of MCL | dMCL | dMCL | 3 | $X_7$ | [0.55-1.05] | $X_{19}$ | [40-180] |
| Anterolateral Bundle of PCL | PCL | PCL_AL | 2 | $X_8$ | [0.85-1.15] | $X_{20}$ | [30-100] |
| Posteromedial Bundle of PCL | PCL | PCL_PM | 2 | $X_9$ | [0.85-1.25] | $X_{21}$ | [30-100] |
| Anterolateral Structure | ALS | ALS | 2 | $X_{10}$ | [0.75-1.25] | $X_{22}$ | [20-125] |
| Popliteofibular Ligament | PFL | PFL | 3 | $X_{11}$ | [0.85-1.15] | $X_{23}$ | [10-90] |
| Posterior Oblique Ligament | POL | POL | 2 | $X_{12}$ | [0.75-1.15] | $X_{24}$ | [30-95] |
| Medial Posterior Capsule | PCAP | PCAP_M | 3 | $X_{13}$ | [0.85-1.25] | $X_{25}$ | [50-100] |
| Lateral Posterior Capsule | PCAP | PCAP_L | 3 | $X_{14}$ | [0.85-1.25] | $X_{26}$ | [50-100] |

Table 4: Optimized ligament parameters for KLA and RKS models for both specimens, respectively. KLA is model calibration performed from data from the knee laxity apparatus. RKS is model calibration performed from data from the robotic knee simulator. $X_1$ represents the first design variable, $X_2$ represents the second design variable, and so on.

| Material Parameter | Design Variable | Ligament | Specimen 1 KLA | Specimen 1 RKS | Specimen 2 KLA | Specimen 2 RKS |
|---|---|---|---|---|---|---|
| Reference Strain | X1 | ACL_AM | 1.14 | 1.07 | 1.16 | 1.14 |
| | X2 | ACL_PL | 1.00 | 1.00 | 1.21 | 1.23 |
| | X3 | LCL | 0.94 | 0.96 | 0.76 | 1.06 |
| | X4 | sMCL_A | 0.79 | 0.80 | 1.05 | 0.94 |
| | X5 | sMCL_M | 0.90 | 0.91 | 1.00 | 0.92 |
| | X6 | sMCL_P | 0.84 | 0.85 | 0.98 | 0.90 |
| | X7 | dMCL | 0.95 | 0.94 | 0.79 | 0.74 |
| | X8 | PCL_AL | 0.91 | 0.92 | 0.88 | 0.89 |
| | X9 | PCL_PM | 0.93 | 0.94 | 0.87 | 0.94 |
| | X10 | ALS | 1.00 | 1.00 | 0.76 | 0.82 |
| | X11 | PFL | 0.99 | 1.14 | 1.15 | 1.15 |
| | X12 | POL | 1.06 | 1.00 | 0.92 | 0.90 |
| | X13 | PCAP_M | 1.25 | 1.25 | 1.08 | 1.11 |
| | X14 | PCAP_L | 1.17 | 1.22 | 1.12 | 1.14 |
| Stiffness (N/mm) | X15 | ACL_AM | 79.41 | 68.53 | 103.85 | 79.21 |
| | X16 | ACL_PL | 114.56 | 115.94 | 50.19 | 58.17 |
| | X17 | LCL | 90.35 | 92.57 | 111.76 | 125.14 |
| | X18 | sMCL | 95.58 | 102.06 | 85.05 | 40.65 |
| | X19 | dMCL | 157.53 | 153.26 | 111.79 | 133.03 |
| | X20 | PCL_AL | 72.87 | 77.64 | 64.25 | 63.70 |
| | X21 | PCL_PM | 82.13 | 85.37 | 30.17 | 30.14 |
| | X22 | ALS | 120.16 | 114.86 | 53.58 | 49.17 |
| | X23 | PFL | 19.90 | 32.56 | 89.52 | 65.68 |
| | X24 | POL | 73.99 | 65.06 | 71.60 | 66.43 |
| | X25 | PCAP_M | 71.12 | 74.38 | 72.41 | 71.49 |
| | X26 | PCAP_L | 61.58 | 60.51 | 70.00 | 67.23 |

Table 5: Predicted ligament loads from 133N anterior and posterior tibial load at three knee flexion angles for KLA vs. RKS models for both specimens. KLA is model calibration performed from data from the knee laxity apparatus. RKS is model calibration performed from data from the robotic knee simulator. All loads recorded under 5 N are represented as a “-“.

| Laxity Direction | Knee Angle (deg) | Specimen | Model | ACL (N) | ALS (N) | LCL (N) | MCL_S (N) | MCL_D (N) | POL (N) | PCAP (N) | PCL (N) | PFL (N) |
|---|---|---|---|---|---|---|---|---|---|---|---|---|
| Anterior | 30 | 1 | KLA | 163.4 | - | - | - | 14.1 | - | - | - | - |
| Anterior | 30 | 1 | RKS | 173.1 | - | - | - | 16.5 | - | - | - | - |
| Anterior | 30 | 2 | KLA | 248.5 | - | - | - | - | - | - | - | 83.3 |
| Anterior | 30 | 2 | RKS | 246.9 | - | 60.1 | - | - | - | - | - | 66.6 |
| Anterior | 60 | 1 | KLA | 161.4 | 13.2 | - | - | 21.5 | - | - | - | - |
| Anterior | 60 | 1 | RKS | 146.3 | 38.3 | - | - | 40.2 | - | - | - | - |
| Anterior | 60 | 2 | KLA | 203.8 | - | - | 18.1 | - | - | - | - | 34.0 |
| Anterior | 60 | 2 | RKS | 206.3 | - | 12.4 | - | - | - | - | - | 35.9 |
| Anterior | 90 | 1 | KLA | 120.3 | 58.2 | - | - | 57.6 | - | - | - | - |
| Anterior | 90 | 1 | RKS | 85.2 | 93.5 | - | - | 84.7 | - | - | - | - |
| Anterior | 90 | 2 | KLA | 194.0 | 16.9 | - | 32.3 | - | - | - | - | 13.0 |
| Anterior | 90 | 2 | RKS | 195.0 | 8.9 | - | 23 | - | - | - | - | - |
| Posterior | 30 | 1 | KLA | - | - | - | - | - | 73.4 | - | - | 89.9 |
| Posterior | 30 | 1 | RKS | - | - | - | - | - | 65.3 | - | - | 75.2 |
| Posterior | 30 | 2 | KLA | - | - | - | 14.1 | - | - | - | 121.8 | 80.6 |
| Posterior | 30 | 2 | RKS | - | - | 33.7 | - | - | - | - | 128.9 | 73.1 |
| Posterior | 60 | 1 | KLA | - | - | - | - | - | 60.0 | - | - | 65.6 |
| Posterior | 60 | 1 | RKS | - | - | - | - | - | 56.1 | - | - | 64.0 |
| Posterior | 60 | 2 | KLA | - | - | - | - | - | - | - | 138.2 | 51.6 |
| Posterior | 60 | 2 | RKS | - | - | - | - | - | - | - | 130.0 | 50.3 |
| Posterior | 90 | 1 | KLA | - | - | - | - | - | 66.4 | 15.2 | - | 58.7 |
| Posterior | 90 | 1 | RKS | - | - | - | - | - | 41.5 | 33.4 | - | 60.2 |
| Posterior | 90 | 2 | KLA | - | - | - | - | - | - | - | 166.8 | - |
| Posterior | 90 | 2 | RKS | - | - | - | - | - | - | - | 155.5 | 32.8 |

Table 6: Predicted GS kinematics during simulated pivot shift at 30 deg of knee flexion with 8 N*m valgus torque and 4 N*m internal torque. KLA is model calibration performed from data from the knee laxity apparatus. RKS is model calibration performed from data from the robotic knee simulator. Models with an "*" denote a simulation with a predicted dislocation between the femur and tibia wherein the kinematics reported may be unreliable.

| Specimen | ACL Condition | Model | F(+)/E (deg) | Vr/Vl(+) (deg) | I/E(+) (deg) | M/L(+) (mm) | A(+)/P (mm) | S(+)/I (mm) |
|---|---|---|---|---|---|---|---|---|
| 1 | Intact | KLA | 30.9 | 2.7 | -13.3 | 0.4 | 3.9 | -17.3 |
| 1 | Intact | RKS | 30.9 | 5.3 | -15.0 | -0.2 | 4.7 | -18.3 |
| 1 | No ACL | KLA | 30.9 | 5.9 | -11.3 | -0.7 | 8.1 | -18.4 |
| 1 | No ACL | RKS | 30.9 | 6.8 | -13.6 | -0.1 | 7.0 | -18.9 |
| 2 | Intact | KLA | 29.9 | -3.1 | -28.8 | 0.0 | 4.6 | -28.8 |
| 2 | Intact | RKS | 29.9 | 1.6 | -31.6 | -2.2 | 1.8 | -30.5 |
| 2 | No ACL | KLA* | 29.9 | 3.7 | -25.1 | -2.1 | 21.6 | -24.4 |
| 2 | No ACL | RKS* | 29.9 | 12.2 | -19.8 | -11.4 | 23.1 | -24.8 |

Table 7: Predicted ligament loads during simulated pivot shift at 30 deg of knee flexion with 8 N*m valgus torque and 4 N*m internal torque. KLA is model calibration performed from data from the knee laxity apparatus. RKS is model calibration performed from data from the robotic knee simulator. Models with an "*" denote a simulation with a predicted dislocation between the femur and tibia wherein the kinematics reported may be unreliable.

| Specimen | ACL Condition | Model | ACL (N) | ALS (N) | LCL (N) | MCL_S (N) | MCL_D (N) | POL (N) | PCAP (N) | PCL (N) | PFL (N) |
|---|---|---|---|---|---|---|---|---|---|---|---|
| 1 | Intact | KLA | 46.9 | 121.8 | 0.0 | 0.0 | 0.0 | 103.1 | 0.0 | 0.0 | 0.0 |
| 1 | Intact | RKS | 17.5 | 145.9 | 0.0 | 0.0 | 0.0 | 86.2 | 0.0 | 0.0 | 0.0 |
| 1 | No ACL | KLA | 0.0 | 163.7 | 0.0 | 0.0 | 0.9 | 85.6 | 0.0 | 0.0 | 0.0 |
| 1 | No ACL | RKS | 0.0 | 159.3 | 0.0 | 0.0 | 5.2 | 81.3 | 0.0 | 0.0 | 0.0 |
| 2 | Intact | KLA | 175.1 | 0.3 | 0.0 | 215.4 | 0.0 | 0.0 | 0.0 | 0.0 | 93.1 |
| 2 | Intact | RKS | 141.6 | 28.7 | 190.2 | 88.6 | 0.0 | 0.0 | 6.9 | 95.7 | 0.4 |
| 2 | No ACL | KLA* | 0.0 | 187.3 | 0.0 | 148.0 | 0.0 | 0.0 | 29.1 | 0.0 | 46.9 |
| 2 | No ACL | RKS* | 0.0 | 271.3 | 41.1 | 184.5 | 0.1 | 0.0 | 0.6 | -0.1 | 0.1 |

FIGURES

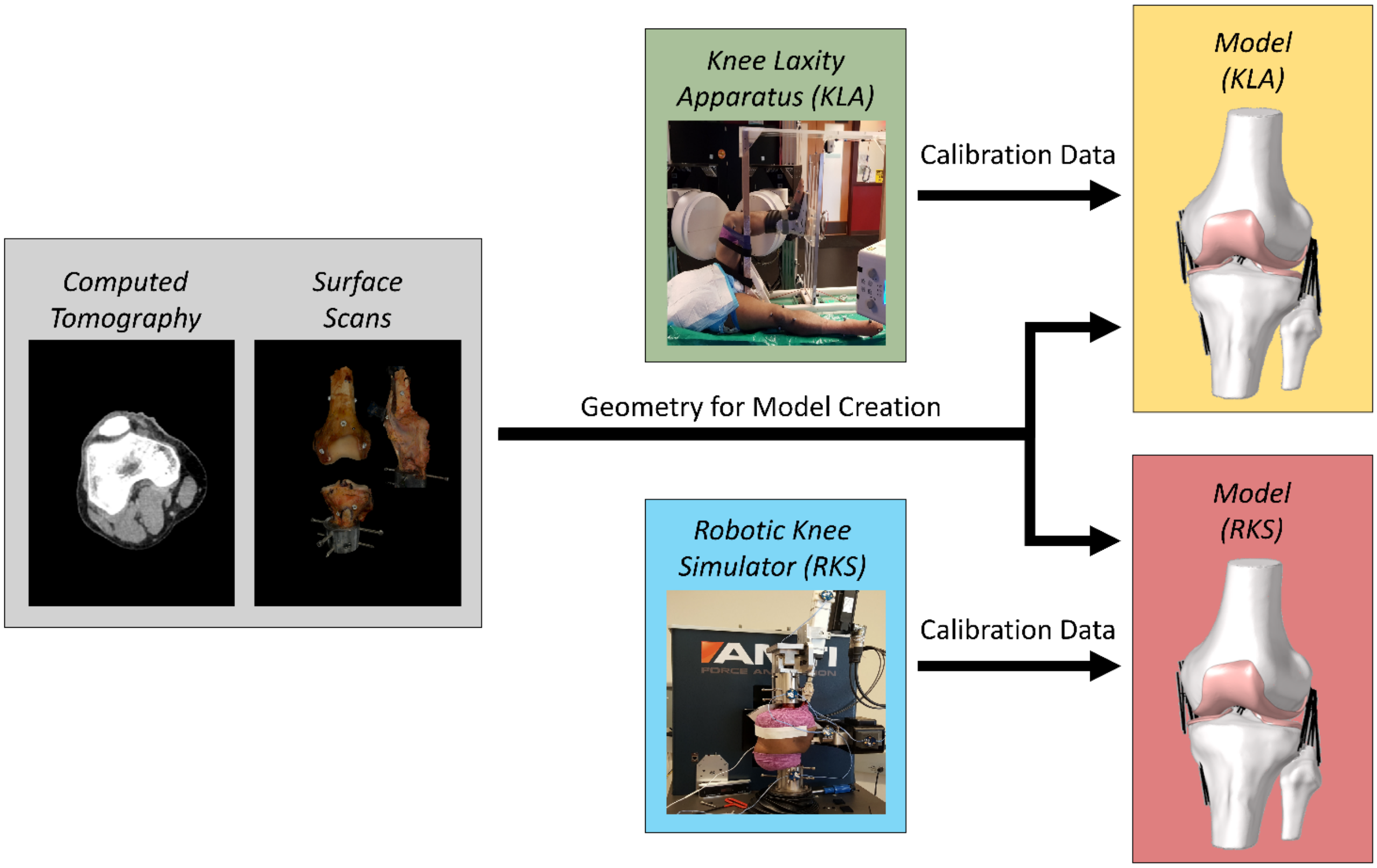

Figure 1: Modeling workflow for each knee specimen. Blue regions are the original imaging data used to create models of the two knee specimens. Green regions are the source of experimental laxity measurements used for calibration. The combined CT scans and surface scans were used to create model geometries for both specimens. Models were then calibrated to the laxity measurements from the knee laxity apparatus and to the measurements from the robotic knee simulator, these models were known as the “KLA” and the “RKS” models, respectively.

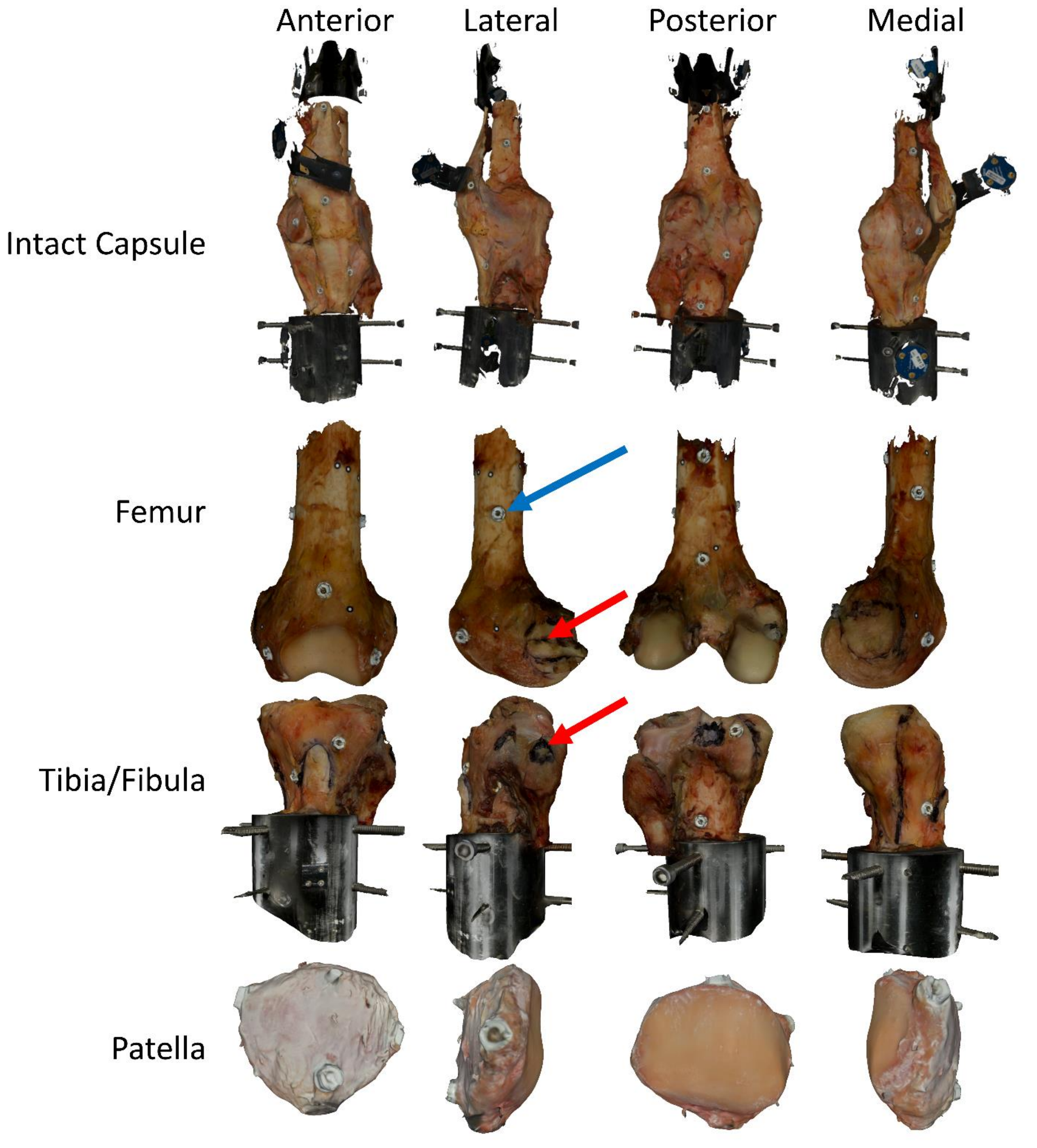

Figure 2: Surface scans of knee at various stages of dissection, intact with capsule, femur only, tibia/fibula only, and patella only. The black highlighted regions on the bones represent different ligament attachment sites identified during the dissection and marked on the specimens using a permanent marker. Bones include fiducial screws and dots to allow for improved registration after the fact, and combining of the original surface data collected from the scanner. Red arrows highlight one of the attachment sites (LCL) identified during dissection. Blue arrow highlights one of the fiducial screws used for registration.

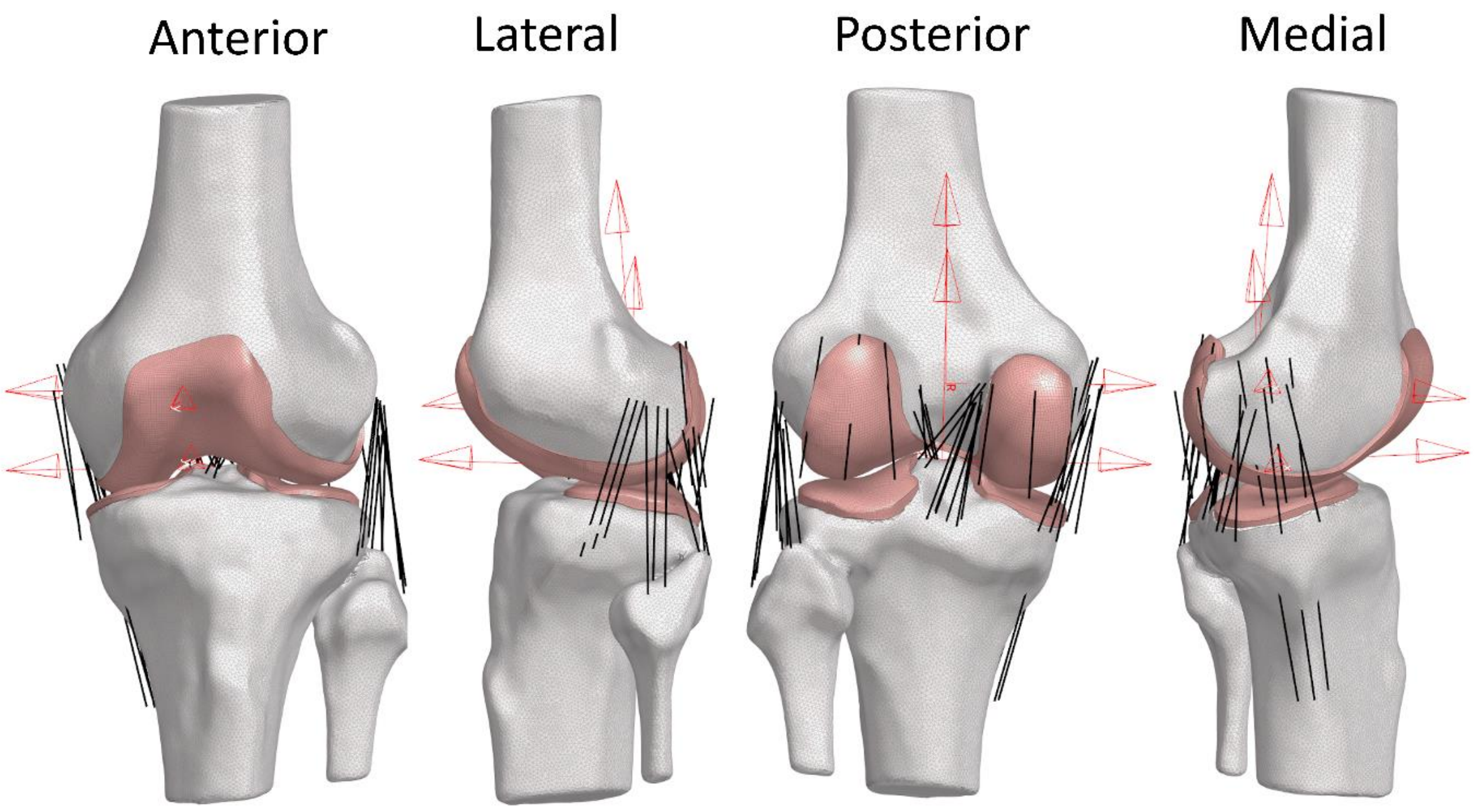

Figure 3: Views of Specimen 1 FEA model. (Red) Femoral and tibial coordinate system definitions using transepicondylar axis (TEA) for the femur coordinate system (White) 2D bone elements (Pink) 3D cartilage elements (Black) 1D ligaments

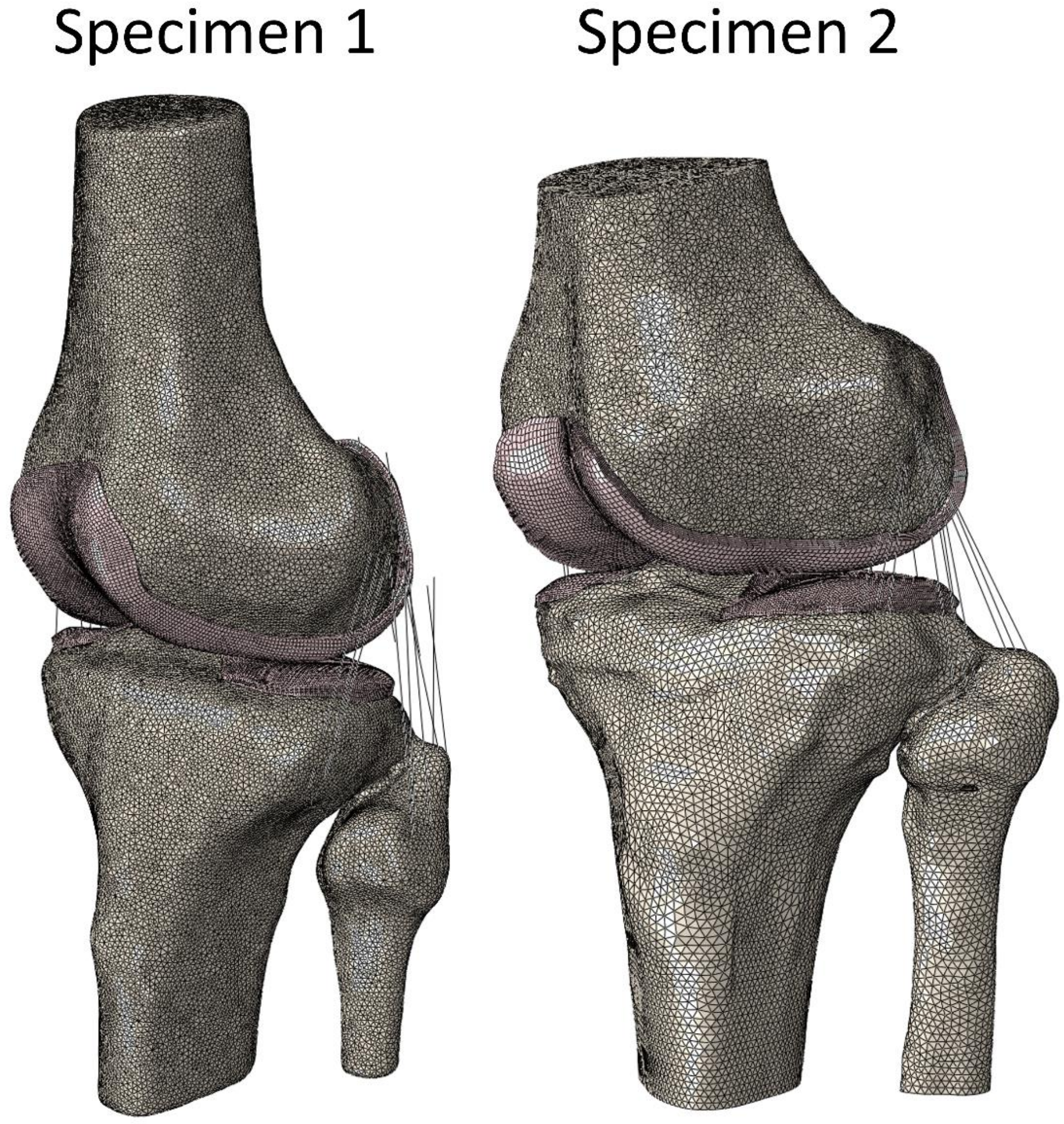

Figure 4: FEA Models of Specimen1 and Specimen 2. (White) 2D bone elements (Pink) 3D cartilage elements (Black) 1D ligaments

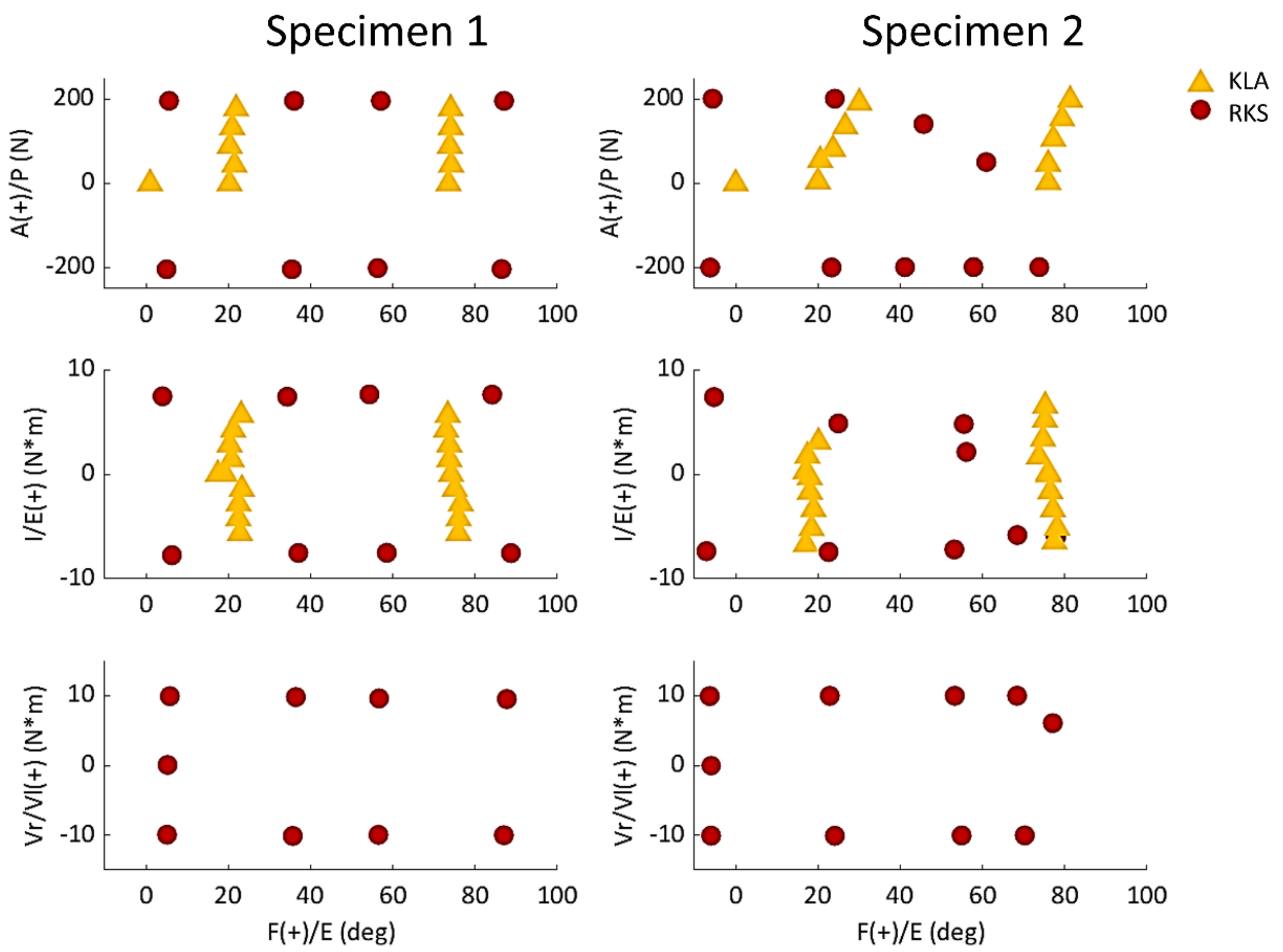

Figure 5: Chosen laxity targets for knee calibration selected from measurements made with the KLA and RKS. Points are the loads and corresponding knee angles of the true experimental data used as targets for the model calibration instead of the approximate knee flexion angle and loads.

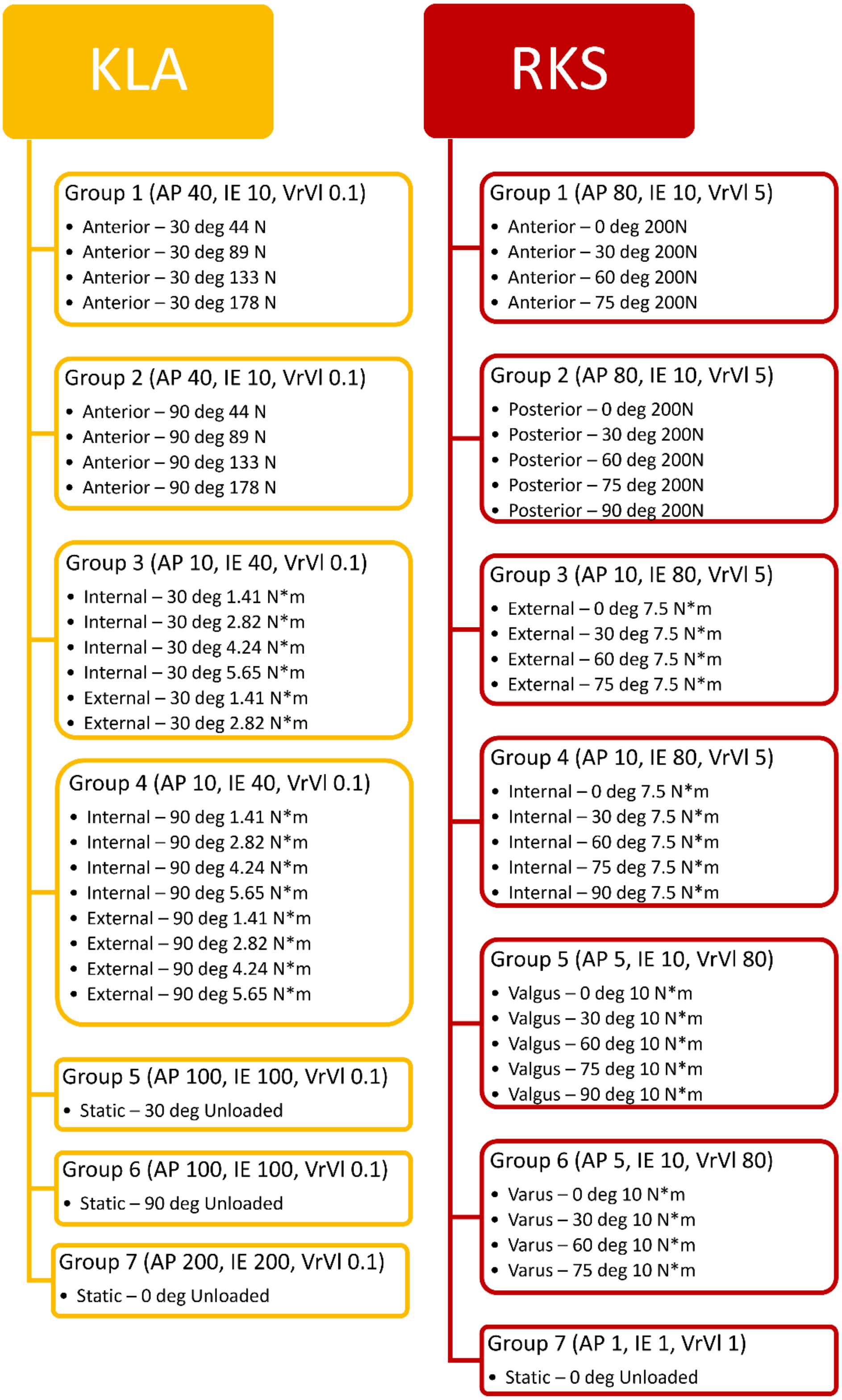

Figure 6: Workflow of the corresponding laxity groups created by grouping together similar trials and experimental laxity points. Values in parentheses represent the corresponding weight applied to errors between the simulation predicted kinematics and the actual experimentally observed kinematics for that DOF.

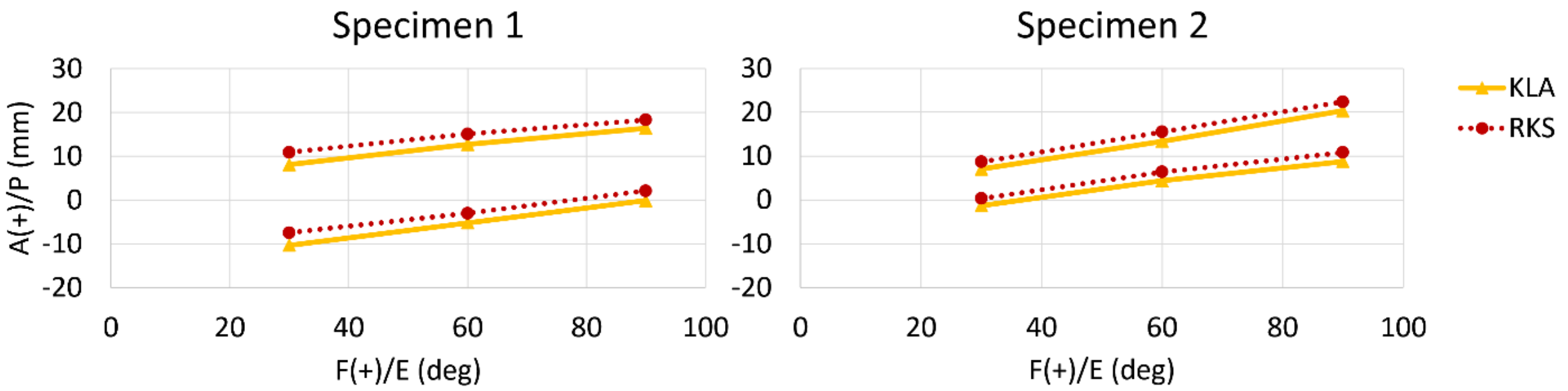

Figure 7: AP GS kinematics from 133N anterior (Top lines) and posterior (Bottom lines) tibial load at three knee flexion angles for all models and both specimens. (KLA) Simulated AP laxity from model calibrated with data from the knee laxity apparatus. (RKS) Simulated AP laxity from model calibrated with data from the robotic knee simulator.

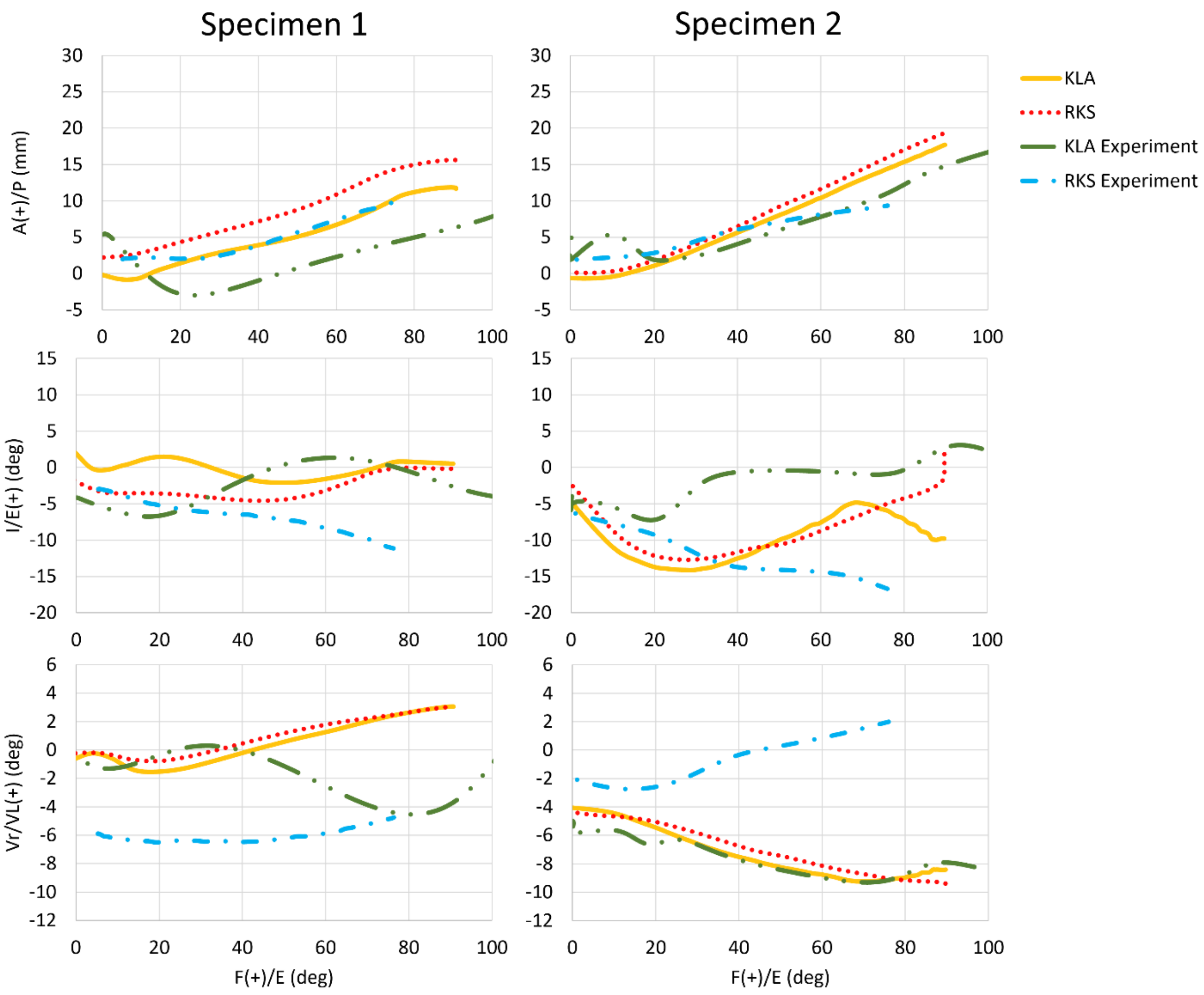

Figure 8: Passive knee flexion GS kinematics for both specimens and all model simulations vs. experimental measures for anterior-posterior (AP), internal-external (IE), and varus-valgus (VrVl) vs. knee flexion angle. (KLA) Simulated passive knee flexion from calibration of model with data from the knee laxity apparatus. (RKS) Simulated passive knee flexion from calibration of model with data from the robotic knee simulator. (KLA Experiment) Experimental passive knee flexion collected via experimenter manual movement through range of motion (RKS Experiment) Experimental passive knee flexion collected via robotic knee simulator no-load motion